\documentclass[twocolumn]{preprint_style}
\leadauthor{Jordan}
\shorttitle{Canalisation of \textit{C. elegans} Development}

\usepackage{caption}
\usepackage{booktabs}

\begin{document}
\setlength{\parindent}{10pt}
\setlength{\parskip}{5pt}

\title{Canalisation and plasticity on the developmental manifold of \textit{Caenorhabditis elegans}}

\author[1,2,\Letter]{David J Jordan}
\author[1,2]{Eric A Miska}

\affil[1]{Wellcome/CRUK Gurdon Institute, University of Cambridge, UK}
\affil[2]{Department of Biochemistry, University of Cambridge, UK}

\maketitle
\begin{abstract}
\textbf{How do the same mechanisms that faithfully regenerate complex developmental programs in spite of environmental and genetic perturbations also permit responsiveness to environmental signals, adaptation, and genetic evolution? Using the nematode \textit{Caenorhabditis elegans} as a model, we explore the phenotypic space of growth and development in various genetic and environmental contexts.  Our data are growth curves and developmental parameters obtained by automated microscopy. Using these, we show that among the traits that make up the developmental space, correlations within a particular context are predictive of correlations among different contexts. Further we find that the developmental variability of this animal can be captured on a relatively low dimensional \textit{phenotypic manifold} and that on this manifold, genetic and environmental contributions to plasticity can be deconvolved independently.  Our perspective offers a new way of understanding the relationship between robustness and flexibility in complex systems, suggesting that projection and concentration of dimension can naturally align these forces as complementary rather than competing.}  
\end {abstract}
\vspace{0.5mm}

\corrauthor{dj333 (at) cam.ac.uk}

\section*{Introduction}
Biological systems are remarkable for their ability to generate reproducible macroscopic dynamics from the complex interactions of large numbers of microscopic components. For example, in animals, the development of an entire organism from a single cell proceeds faithfully each generation even in the presence of environmental fluctuations and molecular noise. Such robustness arises at many spatial and temporal scales e.g., gene expression patterns give rise to reproducible cell differentiation \cite{Rulands2017-us}, neural and muscular activity generate locomotion \cite{Stephens2008-mc}, and interactions between individuals of different species give rise to surprisingly reproducible ecological dynamics \cite{Frentz2015-ec}. This robustness is called \textit{canalisation} \cite{Waddington1957-kj} and dynamics that are canalized are said to be \textit{homeorhetic} \cite{Chuang2019-hz}.

While robust, canalised processes nevertheless allow for important variability; stem cell populations generate diverse tissue types, behaviors respond to stimuli and environmental cues, and populations adapt to changing environments. The structure of this macroscopic variability, however, is much more constrained than the variability intrinsic to the microscopic processes that underlie it. Although gene expression determines the dynamics of the cell cycle, variations in these dynamics are largely insensitive to stochastic fluctuations in the levels of the thousands of proteins. Thus, the ways in which these macroscopic phenotypes can vary is relatively “low-dimensional” compared to the ways in which the individual components can vary.

This projection of variations into a lower dimensional space provides a way for biological systems to be both robust and flexible.  Robustness arises because most variations manifest as excitations onto relatively few phenotypic modes, while flexibility is permitted along these modes. As an example, consider how facial diversity can be generated by the combination of relatively few \textit{eigenfaces} \cite{Turk1991-hb}.  Here the eigenfaces are the modes in this system, and varying the weights of each mode can capture many diverse faces.  However, even large variations in these weights are unlikely to produce non-face images.  Low dimensional representations of phenotypic diversity are ubiquitous in living systems. Variations in phenotypes that tend to a stable state during development have been successfully represented in low-dimensional phenotypic spaces called “morpho-spaces”, for example, morphological traits like the arrangement of flowers on a plant \cite{Prusinkiewicz2007-aj}, the shapes of finch beaks \cite{Campas2010-mh}, of coiled sea-shells \cite{Raup1966-za}, and even the shape of the influenza antigen \cite{Smith2004-yp}. Additionally, while the dimensionality of dynamic, time-varying, and responsive phenotypes is more difficult to define rigorously \cite{Bialek2022-fu}, it has been shown in some cases to be low-dimensional. Some recent examples include the crawling behavior of \textit{C. elegans} \cite{Costa2023-zc} and the neuronal dynamics that underlie it \cite{Brennan2019-dv}, the swimming behavior of both eukaryotic (\textit{Tetrahymena thermophila}) \cite{Jordan2013-el} and prokaryotic (\textit{Escherichia coli}) single celled organisms \cite{Pleska2021-vk}, and the transcriptional trajectories of cells during fate determination \cite{Huang2005-uu}. Concentration of dimension may be an intrinsic property of systems where robustness and control of the macroscopic observables are required \cite{Eckmann2021-oe} and may arise from constraints imposed by steady state growth \cite{Kaneko2021-yg,Klumpp2014-lq}, or trade-offs between phenotypic archetypes suited for different tasks \cite{Shoval2012-az}.

Here we propose that the process of canalisation may be superseded by a more general process of phenotypic concentration of dimension.  That is, genetic networks evolve such that most variations will be projected onto a low dimensional manifold, and that this in turn provides both canalisation and phenotypic plasticity. Concentration of dimensions manifests as the co-variation of traits in a high dimensional multi phenotype space. This structure of the co-variation is called \textit{phenotypic integration} \cite{Pigliucci2003-ua} and its geometric properties can be captured by a phenotypic manifold, i.e., a lower dimensional space that faithfully captures most of the variation observed in a higher dimensional space. The process of finding such a manifold is called dimensionality reduction. The determination of this space and how it is measured can be essential to finding the dominant modes of phenotypic variation, and thus for assessing the degree of canalization. A recent example comes from the developmental program of the wing of the fruit-fly \textit{Drosophila melanogaster} \cite{Alba2021-dr}, which used landmark free morphometrics to uncover a dominant mode of variation that was not apparent from traditional landmark based techniques. 

In this work, we use a custom built automated imaging system to map the phenotypic space of growth and development of \textit{C. elegans}. Using this system, we recorded 673 individual growth curves during the $\approx70$ hours of their development from eggs to reproductive adults, and manually recording the timing of egg hatching and reproductive maturation in single animals. To construct as complete a manifold as possible, we sample from extant variations in different genetic and environmental contexts, including both natural genetic variation and single gene mutants. While the most unbiased approach to sampling genetic variation would be random mutagenesis, because of the immense sampling space, efficient sampling cannot be done. For this reason, we chose to sample ``wild isolates'' of \textit{C. elegans}, i.e. strains collected from nature, and whose collections of genetic changes have been subjected to natural selection. For environmental diversity, we chose to alter the animal's diet; \textit{C. elegans} feed on bacteria, and we chose a collection of bacteria that included the standard laboratory bacteria \textit{E. coli} as well as bacteria isolated from natural sites where \textit{C. elegans} were also collected \cite{Frezal2015-eb}. Measurements were collected for three \textit{C. elegans} wild-isolates, each fed on one of four different bacterial diets, and two mutants of the \textit{C. elegans} laboratory strain N2. 

Using these data, we demonstrate that the space of developmental trajectories can be captured by a low-dimensional manifold and show a correspondence between the directions of fluctuations in a fixed context the directions in which populations will shift due to genetic or environmental changes. We find that the manifold obtained using nonlinear dimensionality reduction techniques captures developmental variations in a way that allows one to neatly decompose the contributions of genetics and environment independently, with the major mode of variation corresponding to environmental shifts and the second mode corresponding to genetic changes. 

\begin{figure*}
    \includegraphics[width=0.99 \textwidth]{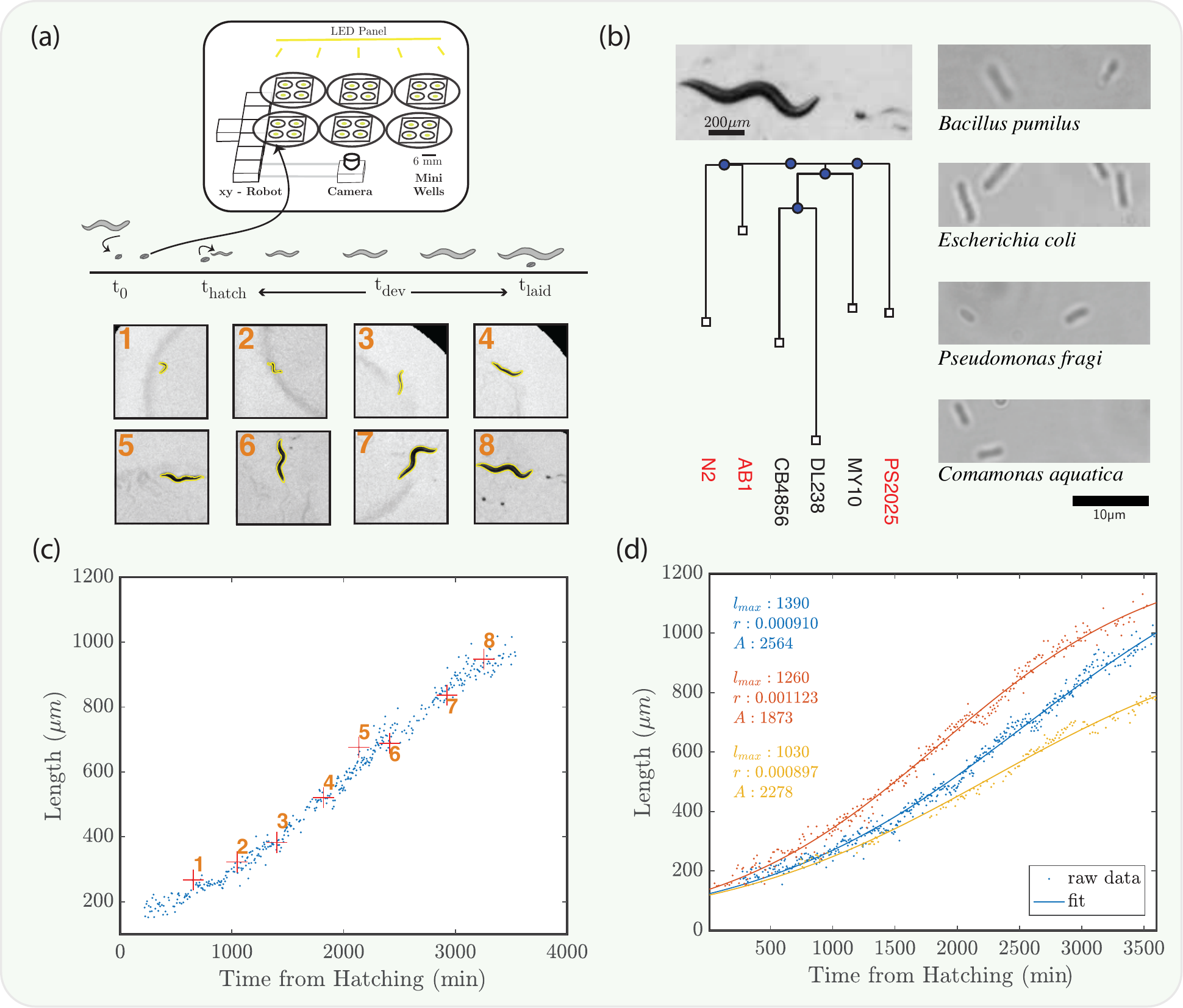}
    \caption{
    \textbf{(a)}  A schematic of the imaging apparatus. Synchronized eggs are transferred into the mini wells at $t_0$.  Hatching $t_{hatch}$ and egg-laying time $t_{laid}$ are manually recorded and give development time $t_{dev}$.  Example images are shown for 10 time points (a, lower), with the computed outline of the worm shown (yellow).  \textbf{(b)} An example image an adult \textit{C. elegans} and an egg (scale bar $200\mu m$). A phylogenetic tree pruned from CeNDR \cite{Cook2017-hq} database of some common wild isolates as well as the three strains used in this work, marked in red. In addition, micro-graphs of the bacteria used as food sources imaged at 100x magnification (scale bar $10\mu m$). \textbf{(c)} The full developmental time course of an animal from hatching through adulthood. The blue points show calculated length over time, with 10 specific points highlighted (red plus) corresponding to each of the images in (a, lower). \textbf{(d)} shows the length measurements and the computed logistic best fit curves from three example time courses.  The parameters of each logistic fit are shown to the left, with each color corresponding to the relevant data and curve. 
    }
    \label{fig:intro_fig}
\end{figure*}

\section*{Results}

Characteristics that change over time are more difficult to quantify and compare than those that are static or near a steady state. Often, dynamic phenotypes such as these are measured and compared at a single fixed time, but in this case, proper alignment or synchronization can be challenging. Ideally, one would like to compare the time series of time-varying phenotypes and compare these directly. The growth of a multi-cellular organism is a good example of such a time-dependent phenotype. Ideally, one would like to compare growth curves aligned to an unambiguous starting time.

To this end, we designed and constructed a low-cost parallel imaging platform capable of measuring \textit{C. elegans} growth for 60 individual animals simultaneously over the course of their $\approx70$ hour development at a temporal resolution of ~0.001 Hz, resulting in a time series of $\approx 200$ observations per animal. In addition to length and area measured automatically, egg hatching, and first egg-laying by mature adults are manually recorded. These not only provide estimates of the animals reproductive development, but also provide a fixed time to which the time series can be aligned.  Reproductive development is in general correlated with growth but need not necessarily be. Because animals grow from approximately 0.2 mm to 1 mm in length during their lifetime, any wide-field system capable of imaging many isolated worms simultaneously would lack the necessary resolution.  To solve this, we developed a system that uses a fixed lens and USB camera that is programmed to move between 6mm diameter custom made wells using an XY plotting robot. Sample images from the time series are shown (Fig. \ref{fig:intro_fig}a) (inset, 1-8), along with the associated time series and the best fit logistic function of the form:

\begin{equation}
l(t) = \frac{l_{max}}{1+e^{-r(t-A)}}
\label{eq:logistic}
\end{equation}

as determined by non-linear least squares fitting, giving three parameters for each curve (Fig. \ref{fig:intro_fig}d). We used this system to record development in a collection of \textit{C. elegans} isolated from the wild and fed on various bacteria, some of which were collected from sites where \textit{C. elegans} were found.  In total, we assayed five unique \textit{C. elegans} genotypes fed on four different bacterial diets (Fig. \ref{fig:intro_fig}b) and collected a total of 673 growth curves, in addition to developmental data, across these conditions.  The developmental data consist of the duration of \textit{ex-utero} embryonic development and the duration of reproductive maturation. These are measured as the time from egg-laying to egg-hatching $t_{hatch}$, and the time from hatching until the animal grows to reproductive age and lays its first egg  $t_{dev}=(t_{laid}-t_{hatch})$ (Fig. \ref{fig:intro_fig}a). While these durations were single scalar quantities, the growth curves consisted of hundreds of points.  Dimensionality reduction could have been performed directly on these high-dimensional vectors after appropriate regularization, but instead these growth curves were fit with the logistic function (Eq. \ref{eq:logistic}), which performed similarly well and whose parameters are easier to interpret (See Supplemental Information).  The three parameters, which we call the maximum length $l_{max}$, the growth rate $r$, and the shift $A$, determine the saturation value of the growth curve, the growth rate at the inflection point, and the temporal shift of the inflection point, respectively. 

\begin{figure}
    \includegraphics[width = 0.97\columnwidth]{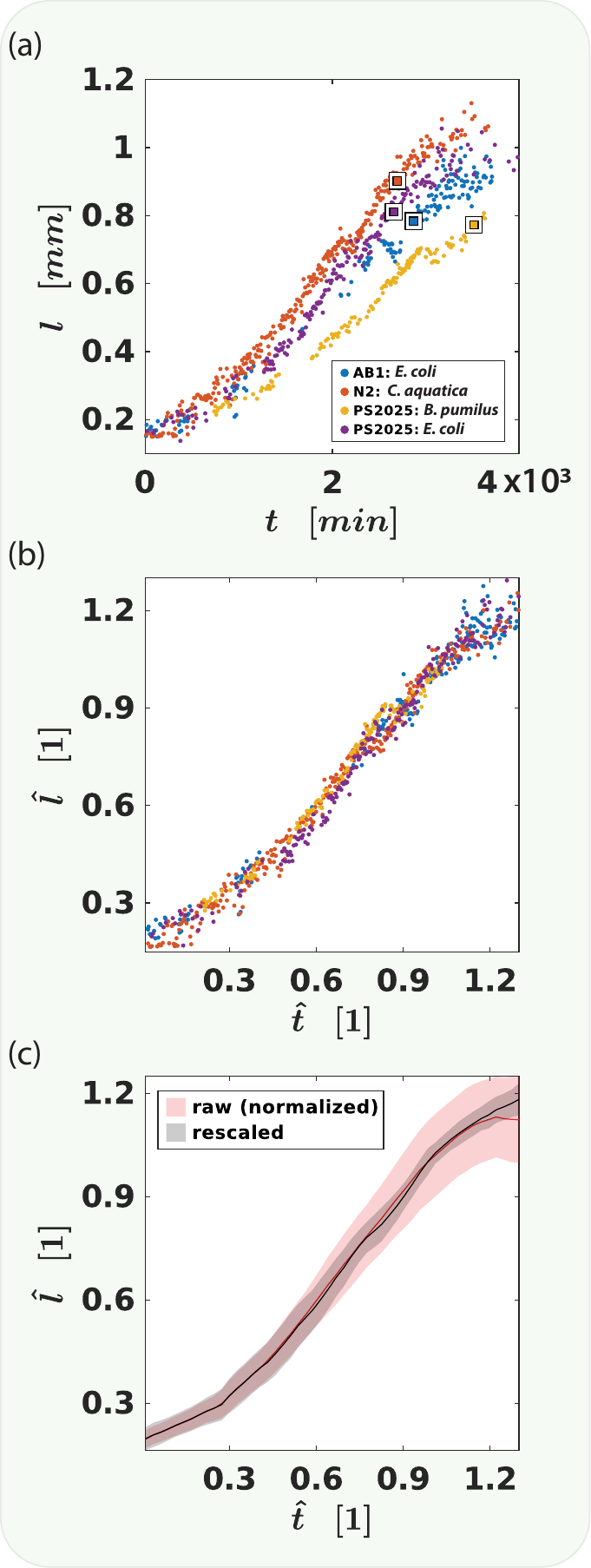}
    \caption{\textbf{(a)} Developmental time course of selected individuals where the square indicates $t_{dev}$ and $l_{dev}$}
\end{figure}

\begin{figure}
    \addtocounter{figure}{-1}
    \caption{
    (continued) for each.  The data can be re-scaled to plot length $l$ as a fraction of development length $\hat{l} = l/l_{dev}$ and time $t$ as a fraction of development time $\hat{t} = t/t_{dev}$ as shown in \textbf{(b)}. Re-scaling the data in this way changes the fit parameters of each logistic function. $[l_{max}, r, A] \to [l_{max}/l_{dev},  r\cdot t_{dev}, A/t_{dev}]$.  The re-scaling of the curves in \textbf{(b)} appears to collapse the growth curves.  This seems to be true in general, as the growth curves re-scaled by their individual $t_{dev}$ and $l_{dev}$ show a smaller standard deviation (\textbf{(c)} black) than the raw data (\textbf{(c)} red) (which is normalized to the average $t_{dev}$ and $l_{dev}$ to facilitate plotting on the same scale).
        \label{fig:rescaling}
    }
\end{figure}

For each growth curve we have an independent measurement of the reproductive development time $t_{dev}$ and of the animals length at this time $l(t_{laid}) = l_{dev}$  (Fig. \ref{fig:rescaling}a, squares).  We can use these additional parameters to rescale each growth curve.  First, we divide the length as a function of time by $l_{dev}$, yielding:

\[\hat{l}(t) = \dfrac{l_{max}/l_{dev}}{1+e^{-r(t-A)}}\]

Then, rescaling time $\hat{t}\to t/t_{dev}$

\begin{equation} 
\hat{l}(\hat{t}) = \dfrac{l_{max}/l_{dev}}{1+e^{-rt_{dev}(\hat{t}-{A}/{t_{dev}})}} 
\label{eq:rescaled_logsitic}
\end{equation}

Thus, the normalization simply rescales the fit parameters from $[l_{max}, r, A] \to [l_{max}/l_{dev},  r\cdot t_{dev}, A/t_{dev}]$
yielding unit-less quantities for the fit parameters. The fitting procedure on the normalized curves recovers the rescaled fitting parameters from the unnormalized data (See Supplementary Information). Interestingly, we find that rescaling the curves to an independently measured parameter, the reproductive developmental duration, seems to collapse the growth curves, consistent with temporal scaling observed previously \cite{Filina2022-vq}, e.g. (Fig. \ref{fig:rescaling}b). 
To compare the variance of all of the rescaled growth curves  to the variance of the un rescaled growth curves, the raw data was normalized to the mean developmental duration and the mean and standard deviation of the resulting growth curves were compared, showing that rescaled growth curves ``collapse'' and that their resulting standard deviation is smaller than for the raw data normalized (Fig. \ref{fig:rescaling}c).

\begin{figure*}
    \includegraphics[width=0.99 \textwidth]{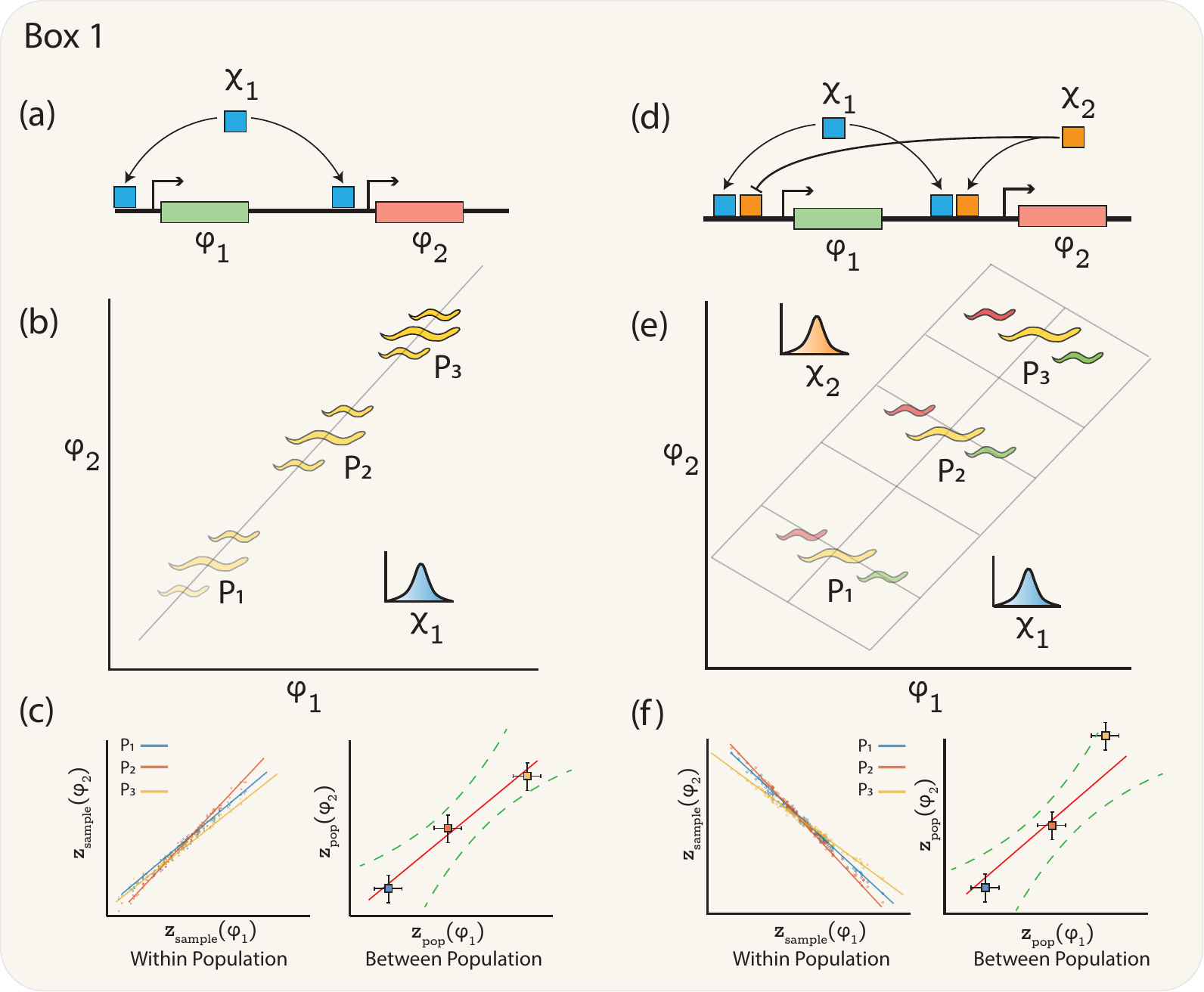}
    \captionsetup{labelformat=empty}
    \caption{
    Box 1: Illustration of relationships between genetic architecture, dimensionality, and the correlation structure of traits.  \textbf{(a)} A simple genetic circuit where green $\phi_1$ and red $\phi_2$ phenotypes are controlled by a single transcription factor $\chi_1$. \textbf{(b)} Different populations (P1-P3) have different mean expression of $\chi_1$ due to either genetic or environmental changes, changing the brightness of the yellow color. Fluctuations in $\chi_1$ around these mean values lead to correlated noise within populations \textbf{(c, left)}. The mean values of $\phi_1$ and $\phi_2$ are also correlated between populations \textbf{(c, right)} , in the same way.  This leads to a one dimensional phenotypic manifold \textbf{(b, grey line)}, which is consistent with the single knob $\chi_1$ In \textbf{(d)}, however, there are two independent transcription factors $(\chi_1,\chi_2)$ that control $\phi_1$ and $\phi_2$.  In each population P1-P3 in \textbf{(e)}, $\chi_1$ sets the mean value of both $\phi_1$ and $\phi_2$ in a correlated manner, changing the overall brightness.  However, fluctuations in $\chi_2$ introduce anti-correlated noise within populations \textbf{(f, left)}, shifting the spectrum to either green or red.  The between population correlation is dominated by the changes in $\chi_1$, resulting in correlated changes in both $\phi_1$ and $\phi_2$ \textbf{(f, right)}.  The within population anti-correlation expands the dimensionality of the manifold \textbf{(e, grey plane)}. Dimensionality can be expanded also by uncorrelated noise both within and between populations (not illustrated). 
    }
    \label{fig:correlation_box}
\end{figure*}

\begin{figure*}
    \includegraphics[width=0.99 \textwidth]{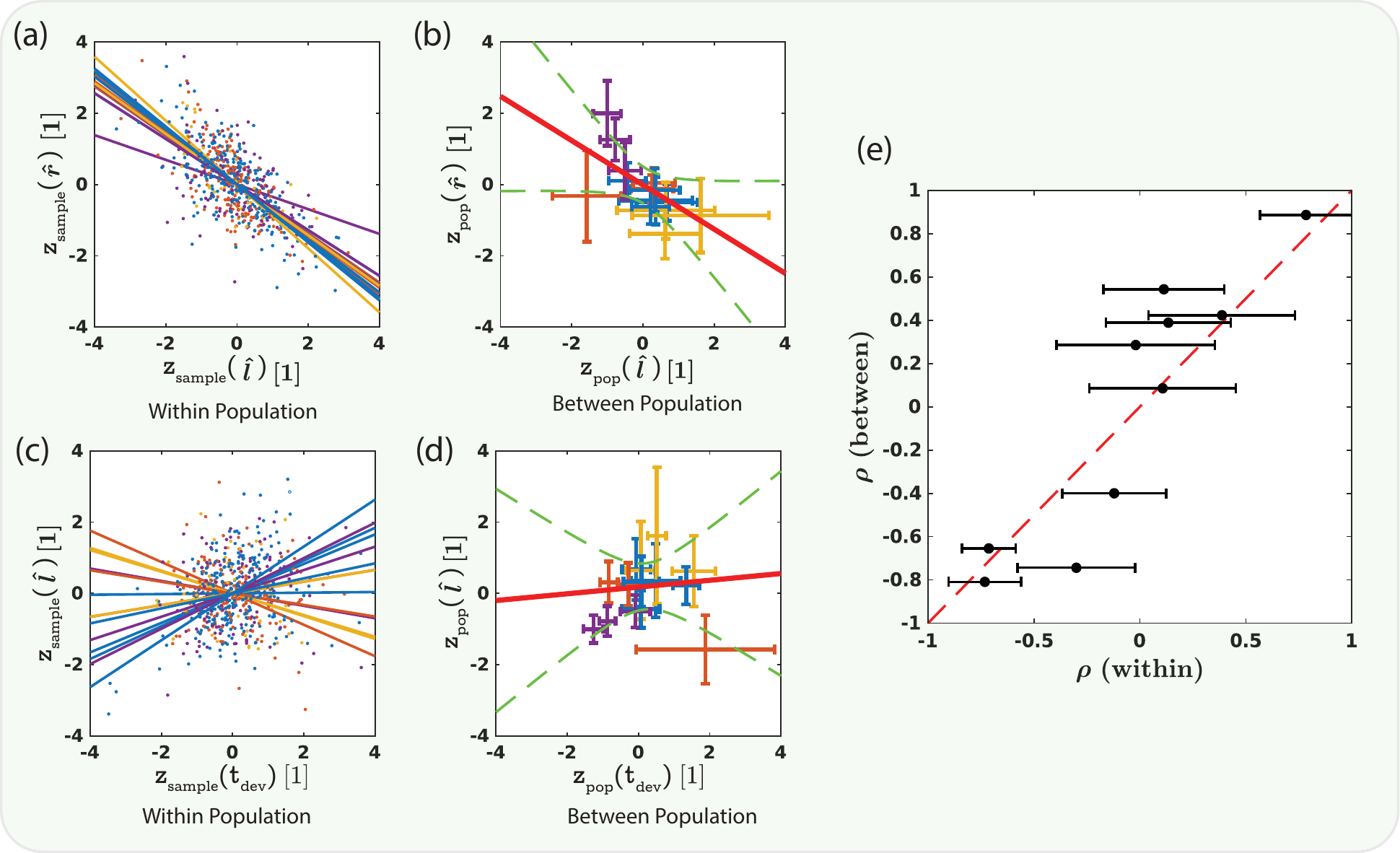}
    \addtocounter{figure}{-1}
    \caption{
     \textbf{(a)} An example showing a pair of developmental parameters $\hat{l}$ and $\hat{r}$ that are correlated within each populations measured.  The scatter plot shows z-score  of  these parameters with respect to each conditions mean and variance, along with the best fit linear regression for that condition (colors indicate the food source).  There is a significant within population negative correlation of these parameters in each condition.  \textbf{(b)} The z-score is then calculated with respect to the mean and variance for all conditions taken together.  These are then grouped by condition and the mean and standard deviation of the z-scores of both parameters are plotted (colors indicate the food source).  The means are also correlated between conditions, and indicated by confidence interval (b, green dashed lines) of the slope of the linear regression (b, red line).  For other traits, there is no significant within population correlation between traits, e.g. between the development time $t_{dev}$ and $\hat{l}$  \textbf{(c)}.  Similarly, in \textbf{(d)}, the between population means are also not significantly correlated .  \textbf{(e)} Summarizes the within and between population correlation coefficient for all pairs or developmental parameters, the mean correlation coefficient for each of the within condition groups is plotted with error bars showing the standard deviation. For the correlation among the mean values, there is only a single value thus there are no vertical error bars.  The red dashed line indicates equivalence, showing that the within population correlations are largely but not exclusively, predictive of the between population correlations. 
    }
    \label{fig:correlation_figure}
\end{figure*}

 The growth curves of \textit{C. elegans} of different genetic backgrounds and grown on differ food sources differ in their maximum length, their growth rate, and in the shift, as well in the durations of \textit{ex-utero} development and reproductive development. However, we find that these parameters do not vary independently. For example, some recent work has shown a negative correlation between growth rate and developmental duration in \textit{C. elegans} and we also observe this in our data. 
 They suggest that this may be a mechanism for reducing variability in adult size in \cite{Stojanovski2022-sp}.  If correlations between traits arise from a mechanism for control such as this,  we would expect variation in individuals to be correlated in the same way as variations between populations.  Strikingly, we find that if parameters are correlated among individuals in one context, these correlations are more likely to appear also between populations in different contexts (To see how such correlations may arise, refer to Box 1). For example, there is a strong negative correlation between the maximum length and the growth rate among individuals in all combinations of conditions.  Likewise, the mean values of these parameters among populations grown in different conditions is also negatively correlated (Fig. \ref{fig:correlation_figure} a,b). In contrast, the length and the duration of reproductive development are not strongly correlated among individuals, and neither are they correlated between populations (Fig. \ref{fig:correlation_figure}c,d).  Computing the correlation coefficient ($\rho$) among individuals in a sample in fixed conditions, and plotting it against the value computed between population means among different conditions shows that this seems to be true in general for different pairs of parameters of \textit{C. elegans} development (Fig. \ref{fig:correlation_figure}e). 

\begin{figure*}
    \includegraphics[width=0.98 \textwidth]{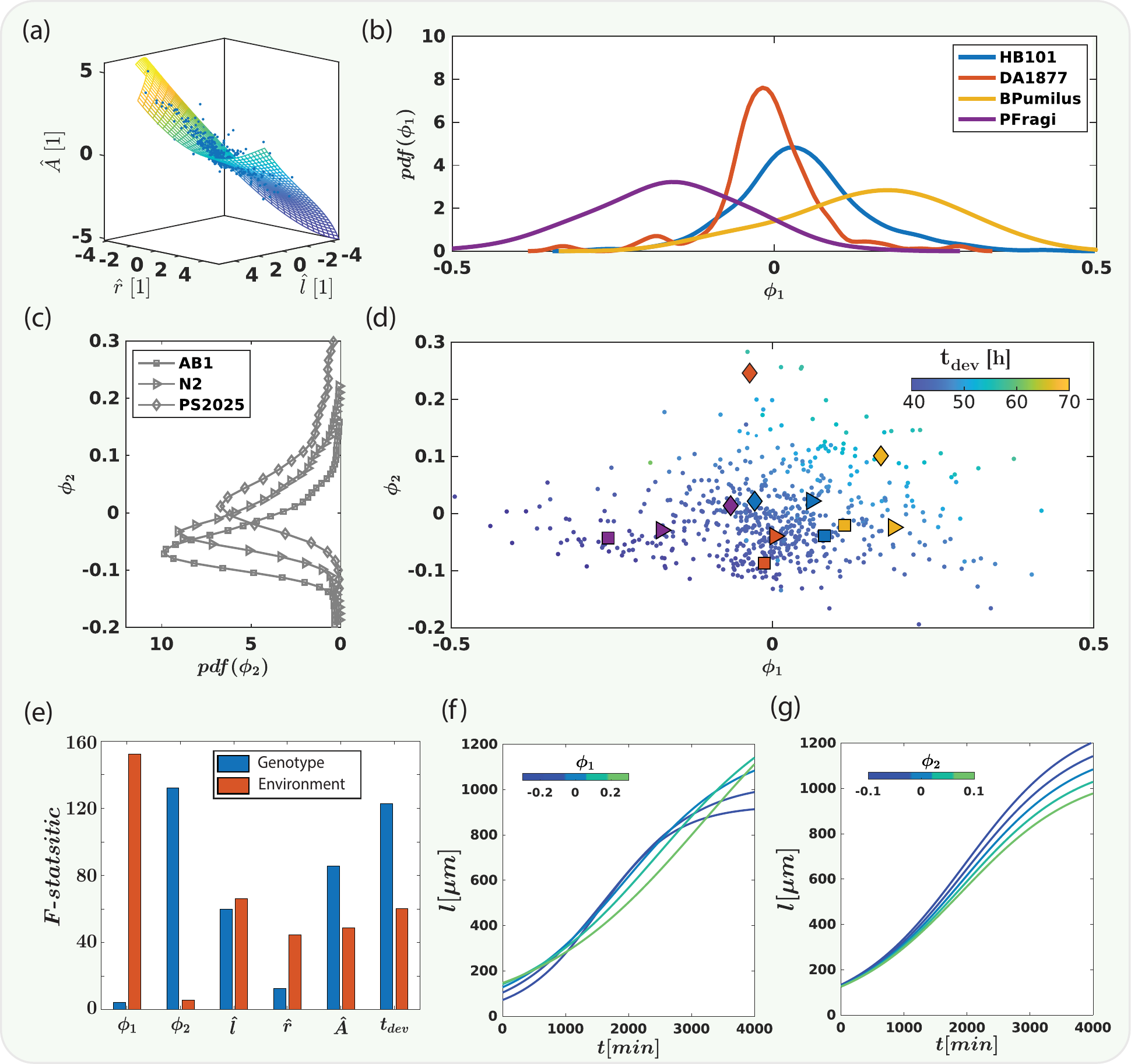}
    \caption{
    \textbf{(a)} The z-score of the three re-scaled logistic fit parameters are shown in a 3-d scatterplot (blue dots). These points lie close to a curved 2-d manifold which was found by performing non-linear principle components analysis (NLPCA).  The flattened manifold is shown in \textbf{(d)} as a scatterplot where the color of the points indicates $t_{dev}$ for each individual. The mean $\phi_1$ and $\phi_2$ for each condition are shown as a combination of symbols (\textit{C. elegans} strain) and color (Bacterial food source). On this manifold, $\phi_1$ seems to separate the environmental conditions as shown in \textbf{(b)} by the marginal distribution over bacterial food sources.  In contrast, the marginal distributions \textbf{(c)} over the \textit{C. elegans} strains shows separation in $\phi_2$. Marginal distributions were computed with a kernel density estimator. This separation in $\phi_1$ and $\phi_2$ can be quantified by computing the f-statistic for a linear regression model on taking genotype and environment as regressors \textbf{(e)}. $\phi_1$ regresses primarily on environment and $\phi_2$ on genotype.  Interestingly, regressing the three logistic fit parameters without first performing NLPCA results in mixed mixed regression on both genotype and environment for each. In \textbf{(f)} to determine the effect of varying $\phi_1$ on the shape of the growth curve, $\phi_2$ was fixed to 0 and $\phi_1$ was varied through a range, as indicated by the colorbar, with coordinates being converted back from the unit-less quantities.  Similarly in \textbf{(g)},$\phi_1$ is fixed and $\phi_2$ varied. 
    }
    \label{fig:manifold_figure}
\end{figure*}

The geometric structure of these correlations can be captured by a low dimensional manifold in the ambient phenotypic space. Non-linear principal components analysis \cite{Scholz2005-xp} was used to determine the shape of this manifold from the ambient space of rescaled logistic parameters with the developmental durations included (Fig. \ref{fig:manifold_figure}a). The first two non-linear principal components capture $93\%$ of the variance. While this dimensionality reduction may seem modest, in fact, the true dimensionality of the growth curve space is higher with some of the dimensions having already been compressed by the logistic fit. Within this embedding of growth curves, animals tend to cluster in $\phi_1$ according to their food source (Fig. \ref{fig:manifold_figure}b) and in $\phi_2$ according to their genetic background (Fig. \ref{fig:manifold_figure}c) (note only natural genetic variants are shown).  This decomposition can be quantified with a linear regression model predicting either the NLPCA parameters $(\phi_1,\phi_2)$ or the rescaled logistic fit parameters $(\hat{l},\hat{r},\hat{A})$ using the genetic or environmental conditions as independent variables.  Linear regression models of the form:
\[ y(G,E) = \beta_0 + \sum_i\beta_{1,i}G + \sum_j\beta_{2,j}E + \epsilon\]
were fit where $\epsilon$, with $y\in(\phi_1,\phi_2,\hat{l},\hat{r},\hat{A})$ , $G$ and $E$ are indicator variables which take the value of 1 or 0 depending of the strain $i$ and food $j$ in that condition, and $\epsilon$ are the residuals to be minimized. In Fig \ref{fig:manifold_figure}b and Fig \ref{fig:manifold_figure}c, the distributions grouped by environment and genotype respectively seem to have different means.  We can quantify this using the F-statistic, which is the variance of the between group means divided  by the mean of the within group variances.  Using this, we can see that grouping the data by environment does in fact give distinct distributions in $\phi_1$ and grouping by genotype gives distinct distributions in $\phi_2$ (Fig. \ref{fig:manifold_figure}e).  However, a linear regression of the parameters before dimensionality reduction does not give a clean separation between genotype and environment.

The eigenfunctions of the non-linear principle components analysis cannot be derived analytically, but can be investigated by varying each component while keeping the others fixed. If $\phi_2$ is fixed $(\phi_2=0)$, the shapes of the resulting growth curves as $\phi_1$ is varied reflect the strong anti-correlation between the parameters $l_{max}$ and $r$. For positive values of $\phi_1$, animals grow quickly during their maximum growth phase but are ultimately shorter (Fig. \ref{fig:manifold_figure}f, blue curve).  In contrast, for negative $\phi_1$, animals grow more slowly at their peak, but are longer as adults (Fig. \ref{fig:manifold_figure}f green curve).  This may indicate a potential trade-off between the speed of growth during development and the ultimate size achieved by adult animals. In contrast, variation along $\phi_2$ result both in growth that is slower and in animals that are shorter as adults (Fig. \ref{fig:manifold_figure}g, green curve) or both faster and longer (Fig. \ref{fig:manifold_figure}g, blue curve).  Interestingly, variation \textit{along} $\phi_1$ does not seem to affect the duration of reproductive development as much as along $\phi_2$ as shown by the color of the plotted points (Fig. \ref{fig:manifold_figure}d). Points away from the boundary in the positive $\phi_2$ direction correspond to slower reproductive development.  

\section*{Discussion}

Biological systems are remarkable in part because they seem to be both incredibly robust and simultaneously flexible.  Biological process faithfully regenerate complex developmental programs every generation, yet these same processes have given rise to an overwhelming diversity of complex matter. Canalisation and developmental plasticity seem to be opposing forces.  One can imagine that these forces ebb and flow, acting in sequence. Cryptic variation which can be suppressed or revealed in certain circumstances, e.g. by the chaperone HSP-90 \cite{Rutherford1998-is}, is a good example of this, and it has been shown that disruption of the chaperone at the molecular level act as the switch from robustness to plasticity, even in complex phenotypes such as the morphology of an animal \cite{Sieriebriennikov2018-jk}.  However, this may be only part of the story.  Canalisation and flexibility may in fact be complementary properties of systems that are organized to generate low dimensional phenotypic manifolds. Concentration of dimensionality is an intrinsic property of complex high-dimensional systems, even if only trivially because of the Johnson-Lindenstrauss lemma \cite{Johnson1984-en}.  However, the emergence of low dimensions in biological systems is often orders of magnitude more than random projection would predict \cite{Eckmann2021-oe}. At the heart of concentration of dimensionality lies projection. We propose that is natural to view robustness and plasticity as analogous to projection of variation onto, or orthogonal to, a low dimensional manifold, respectively. Given a distribution of variations that arise from environmental heterogeneity, stochastic fluctuations, and genetic mutations, whether a given phenotype or set of phenotypes is buffered or responsive will depend on the projection function from high dimensional chemical space to the low dimensional phenotype space.  In cases where a phenotype is highly buffered to some set of variations, almost all those variations will be orthogonal to the phenotypic manifold, and variations for which the phenotype is plastic will have large projections onto the manifold.  In this scenario one might envision the process of evolution as one of shaping the projection function, and thus the resulting phenotypic manifold, given the statistics of distribution of variations.  The process of adaptation can then be viewed as the process of learning the optimal projection over the prior distribution of fluctuations, i.e. an environment to phenotype rather than genotype to phenotype map \cite{Xue2019-wy}, although the environment to phenotype map is certainly a product of the architecture of the gene regulatory networks of the organism.  

\begin{figure*}
    \includegraphics[width=0.99 \textwidth]{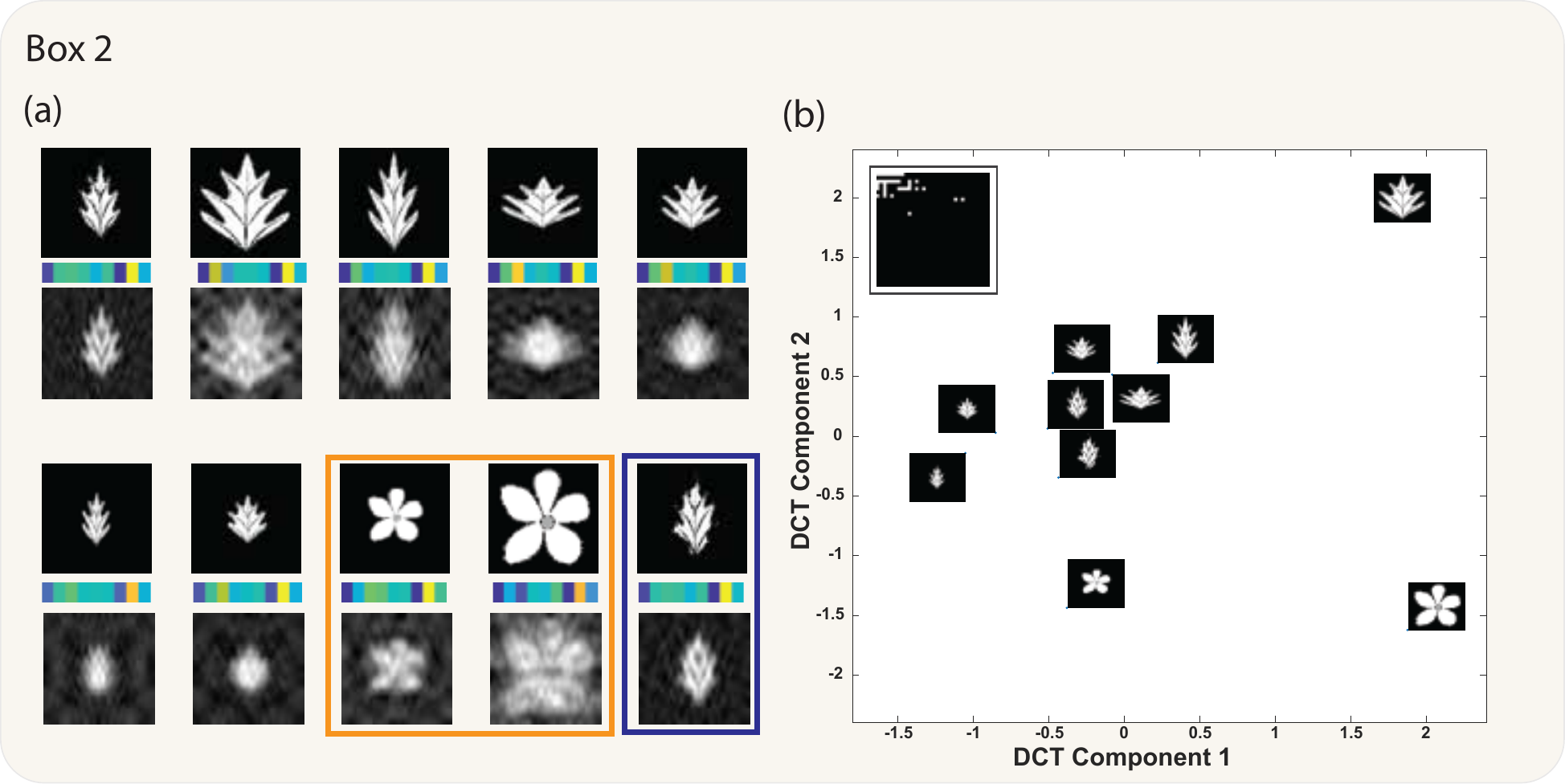}
    \captionsetup{labelformat=empty}
    \caption{
    Box 2: Illustration of robustness and flexibility resulting from projection onto a low dimensional manifold. \textbf{(a)} In this example, 64x64 pixel leaf images represent a high dimensional environment to be sensed.  The discrete cosine transform is then used as a projection operator.  Past evolutionary history has optimized this projection function to recognize the top left leaf in panel \textbf{(a)} and for this, the top 35 modes capture $90\%$ of the energy.  The retained modes are shown in \textbf{(b, inset)}. In \textbf{(a)}, we can see a variety of environmental inputs and their reconstructions from their low dimensional representations, the first 10 mode weightings are shown as color band below each image in \textbf{(a)}.  Even though this projection operator evolved for the leaf in the top left, it reconstructs the different leaf images reasonably well.  In particular, high frequency variation around the original leaf shape is buffered \textbf{(a, blue box)}, resulting in canalisation of the representation such fluctuations in the environmental input.  Furthermore, even distinct environmental inputs such as the flowers in \textbf{(a, orange box)} are reconstructed fairly well, demonstrating that this projection operator is robust, even in very divergent contexts. \textbf{(b)} shows the embedding onto the first two discrete cosine transform components.  From this, we can visualize both the robustness and the plasticity associated with this projection operator.  The leaves cluster separately from the flowers, revealing flexibility, and the two leaves that only differ in high frequency modes cluster together, showing robustness, with the variation between reconstructed representations smaller than that of inputs (See Supplemental Information).  Here the first component is correlated with overall size and the second seems to separate leaves from flowers. It is interesting to note that the projection operator was not optimized in any way to distinguish leave from flowers, and in fact, the eigenfunctions used are the generic ones from the DCT.  Over evolutionary time, organisms could optimize not only which modes are retained to better capture the prior distribution over contexts, but also could optimize the shape of the eigenfunctions themselves.  This panel depicts an environmental sensing mechanism, but equally important, though not shown here, would be the phenotypic execution mechanism.  The structure of the network which maps the low dimensional representation of the environment back into a high dimensional phenotype will also be an important source of variability and affect both robustness and plasticity. 
    }
    \label{fig:projection_box}
\end{figure*}

It is tempting to look for the genetic basis of emergence of low dimensional phenotypes, especially in \textit{C. elegans}, which has many genetic tools available.  However, we believe it is unlikely that a single gene or a genetic module will underlie this process. Rather, it will likely be a property of the entire network of molecular interactions that underlie complex phenotypes. Nevertheless, there may be important connections between the structure of phenotypic manifolds and evolutionary dynamics. It has been proposed that movement along, rather than away from such manifolds constitutes a path of “least resistance” for genetic changes which might fix phenotypic variations. While some evidence supports this hypothesis, e.g. a study of the integration of phenotypic and life-history traits in the flowering plant \textit{Arabidopsis thaliana} \cite{Pigliucci2001-qe}, other evidence from the morpho-space of the greenfinch \textit{Carduelis chloris} suggests that within population correlations may not predict between population correlations \cite{Merila1999-wd}. This discrepancy may be due in part to how traits are quantified, how the dimensionality reduction is performed, and to whether the populations included in the analysis sample natural genetic variation, genetic mutations, environmental variations, or combinations of all three. While the question of when and in what conditions the ``directions'' of phenotypic plasticity are predictive of the directions of subsequent genetic evolution remains open, phenotypic plasticity in canalized traits has been shown to be associated with rapid evolutionary diversification. An example comes from the polyphenism in the feeding structures of nematodes, in which the acquisition of mouth form plasticity is associated with an increase in evolutionary rates. Interestingly, even if only one of the two alternative forms is subsequently fixed, the underlying genetic architecture seems to maintain an expanded phenotypic manifold which facilitates future exploration \cite{Susoy2015-ln}.  While traditional forward and reverse genetics may not be the ideal approach, artificial evolution for expanded phenotypic manifolds seems promising, especially with an automated system that can map individuals to the phenotypic manifold in real time and use this as a selection criterion. 

In physics, statistical mechanics provides tools to connect microscopic and macroscopic dynamics and to describe the behavior of the macroscopic observables that arise from systems with large numbers of identical interacting components. Examples include the Navier-Stokes equation for fluid flow or the Fokker-Planck equation for diffusion. The coarse graining of many interacting degrees of freedom into a few dominant modes can be formulated precisely in terms of projection operators \cite{Mori1965-uy,Zwanzig1961-tc,Nakajima1958-ke}.  These projections often make use of time scale separations. For example, the random motion of a Brownian particle results from its many collisions with surrounding molecules, but these collisions occur very fast and can thus be treated as white noise. For near equilibrium systems, the rigorous relation between the fluctuations arising from the many unobserved degrees of freedom and the evolution of the system’s macroscopic observables are known collectively as Fluctuation-Dissipation relations \cite{Callen1951-zu,Kubo1966-kl}. While biological systems are characteristically far from equilibrium, similar relations have been observed between phenotypic variability and evolutionary response \cite{Sato2003-dx,Tang2021-pz}. In fact, even in systems far from equilibrium, relations of this sort can arise similarly when there is a separation of both observed and unobserved degrees of freedom and of time-scales \cite{Jung2021-bj}. However, the origin of low dimensional dynamics and of fluctuation-dissipation like relationships in biological systems remains a mystery. A recent intriguing proposal for the origin of such relations and the emergence of low dimensionality in biological systems is the evolution of robustness. If the same evolutionary forces that increase robustness to noise also tend to increase robustness to genetic perturbations this could account for both phenomena \cite{Kaneko2021-yg}. Recently, Murugan and colleagues have presented a model describing how mutational perturbations may be constrained by global epistasis to excite only a few slow or \textit{soft} modes with examples from protein elasticity and gene regulatory dynamics \cite{Husain2020-qs}.  Their work shows that such a relationship between mutational induced and physically induced deformations is expected mathematically for protein elasticity. The existence of such slow modes may also be related to the seemingly universal emergence of \textit{stiff} and \textit{sloppy} modes from parameter space compression \cite{Gutenkunst2007-zn,Machta2013-qv}. 

In \textit{C. elegans}, development has been shown to be controlled  by a massive gene expression oscillator that is comprised of of $\approx 3700$ genes \cite{Hendriks2014-nf,Meeuse2020-nt}.  The existence of such an oscillator evokes a direct analogy to projection operator techniques as this network only has a few excitable oscillatory modes similar to a Fourier transform or the cosine transform employed in Box 2. Recently, the components of a gene regulatory network that comprises a central clock to control this oscillator has been identified \cite{Meeuse2023-vv}.  As a follow up to this work, we would like to assess how mutations in these core components affect how fluctuations in gene expression are projected onto the main oscillatory modes that control molting, and how this in turn manifests on the developmental manifold of \textit{C. elegans}.  

In this work we have focused on variability resulting from combinations of dietary and genetic perturbations, and we have done so in only fixed environments.  In the future, it will be interesting to perform experiments in which the developmental trajectory was perturbed by some impulse or step function and to observe the resulting relaxation.  \textit{C. elegans} development is highly temperature dependent, and it would be interesting to first assay they developmental trajectories at different temperatures and then to perturb development using various temperature shift protocols.  In addition, if we could find perturbations that tend to generate displacement in specific directions on the phenotypic manifold, we could test quantitatively for fluctuation response type relationships.

The projection operator perspective leads one naturally to consider how organisms might learn their particular projection function.  Evolution by mutation and selection will surely play a large part in this process, especially when an organism must adapt to changes to the prior distribution of variations, but there is no reason, in principle, that this operator cannot be shaped also by within generation ``learning'' and other non-genetic and epi-genetic feedback mechanisms. 

\section*{Conclusions}
Biological systems are remarkably robust, and yet the same mechanisms that underlie this robustness have also generated the incredibly diversity of life on earth. In this work we attempt to reconcile these seemingly opposing forces by studying the structure of variability in the development of an animal in the contexts of different genetic backgrounds and environments.  We have used high resolution imaging data of the growth of the nematode \textit{C. elegans} in environmental and genetic contexts to map the variability of development. We find that traits which are correlated within a particular context predict whether the mean values of those traits will be correlated among different contexts. Correlation between traits indicate that the true dimensionality of the system may be lower, and we find a parsimonious low dimensional representation of the variability, in which the contributions of genetic variation on the phenotype are separable from those of environmental variation by a simple linear model. We present a framework in which the emergence of low dimensionality provides a basis for both robustness and plasticity. Concentration of dimensionality seems to be an intrinsic property of the chemical and molecular networks that generate living systems, and of complex dynamical systems in general.  While beyond the scope of this work, we hope that the connection between projection operators in physical systems, which explain how reproducible coarse grained dynamics arise from the stochastic influences of innumerable microscopic components, may inform the theory in biology of how reproducible developmental dynamics arise from the interactions of similarly large numbers of bio-molecules.  

\section*{Methods}
\paragraph{Imaging hardware} The imaging hardware consists of a single 3.2 MP monochome camera (Flea3, Teledyne FLIR) mounted to the arm of an XY-plotting robot (Eleksdraw, Eleksmaker) which moves the arm using two stepper motors controlled by a GRBL stepper motor controller that interprets G-code sent via a USB serial connection using the MATLAB (R2018b, Mathworks) fprintf command. Returning to the same position was accurate to with a few hundred microns, and wells were recentered periodically by fitting a circle to an intensity thesholded image of the well and zeroing the offset. Illumination was provided from above by an LED light panel. To maximize contrast, oblique illumination was blocked by a sheet of black acrylic in which an array of square holes was laser cut to allow for the direct illumination to pass though.  A 50mm f/1.4 lens (NMV-50M1, Navitar) was attached to the camera via a 40mm lens tube (CML40, Thorlabs). In this system, one pixel in the image corresponded to $2.67 \mu m$, giving an effective magnification of $0.93x$. Detailed instructions and a parts list with suppliers is available upon request.

\paragraph{Multiwell plates} Multiwell plates were made by gently placing a pre-cut acrylic form into a standard 30mm Falcon petri dish filled with 5ml of liquefied NGM-Gelrite medium.  The multi-well acrylic forms were 24mm square, with a 2x2 grid of 6mm diameter circular wells 4cm from the edge and 10 cm center to center, were cut from 3mm thick black acrylic using an LS 6090 PRO Laser Cutter (HPC Laser Ltd).  After placing the multiwell plate into the molten solid media so that it rested on the surface, the plate was allowed to set and to dry for 1 hour covered at $20^{\circ}$C, and seeded with $2 \mu l$ of bacteria corresponding to approximately 18 million cells, as measured using a Petroff-Hausser Counting Chamber (Hausser Scientific).  

\paragraph{Temperature control}
Plates were kept in an insulated, temperature controlled box during imaging, which itself was in a temperature controlled room set to $20^{\circ}$C.  The actual average temperature in the room was $19.7^{\circ}$C.  Temperatures were measured with a custom thermometer; the signal from a linear thermistor (Omega Engineering) was difference amplified against a known voltage corresponding to $20^{\circ}$C.  The amplified signal (Gain = 2) was recorded in MATLAB (Mathworks) from the analog input of a DAQ (Labjack).  This was fed into a Proportional Integral Differential (PID) control script whose output was a voltage that controlled a Push-Pull current amplifier driving a Peltier effect element (Custom Thermoelectric).  Fans were used to distribute air from the heat sinks on each face of the Peltier element either into the enclosure or as exhaust.  Temperature was maintained to within 70 mK of the set point ($20^{\circ}$C).

 \paragraph{Image processing}
 Images were captured directly into Matlab using the built-in video input object class. Moving objects were extracted from each image using background subtraction, to generate a region of interest with the largest amount of detected motion, as measured by the largest pixel difference.  Within this region of interest, objects were detected by contrast with the background by applying a threshold to a laplacian of gaussian filtered image.  Filtering was performed using the Matlab \textit{imfilter} function with the \textit{fspecial} function with filter size 15x15 and filter standard deviation of 1.5.  Connected components were extracted from this thresholded image and the locations and properties of connected components were recorded from the resulting black and white image.  Area (in pixels), and centroid location were calculated using the MATLAB function \textit{regionprops}, and length was computed using the MATLAB functions \textit{bwmorph} to extract the skeleton, the function \textit{bwgeodesic} to compute the length.  These properties were used to classify each blob as a worm or not, based on the output of a pre-trained support vector machine.  All images were also saved for later inspection, which was used to determine the egg-hatching and egg-laying times manually.  All code used for analysis and data processing is available on Github, including data from processed images (Raw images are available on request).
 
\paragraph{Nematode culture and strains}
\textit{C. elegans} was grown on NGM agar plates and fed \textit{E. coli} HB101 for standard maintenance at 20 °C. Bristol N2 was used as wild-type strain \cite{Brenner1974-se}.  In addition to N2, two wild isolate strains were used, AB1 isolated in Adelaide, Australia, and PS2025 isolated in Pasadena, California.  In additon, two single gene mutants of N2 were used, \textit{sid-2(mj465)},  a mutant that is not competent for RNAi by feeding, and \textit{c28h8.3(mj649)} a catalytic mutant of a putative helicase.

\paragraph{Bacterial culture and strains}
In addition to \textit{E. coli} HB101, three other bacteria were used as food sources.  \textit{Comamonas aquatica} DA1877 was used as it has been previously shown to increase the rate of \textit{C. elegans} development by providing supplemental vitamin B12 \cite{Watson2014-sl,MacNeil2013-ge}.  We also chose \textit{Bacillus pumilus} and \textit{Pseudomonas fragi} from the wild bacteria collection \cite{Frezal2015-eb} as we had previously observed large developmental differences on these food sources. 

\paragraph{Synchronization by coordinated egg laying} To generate populations of synchronized animals without bleaching and starvation, young egg-laying adults (50-75 hours post hatching) were gently picked with a platinum wire to a fresh NGM plate seeded with the appropriate bacteria and allowed to lay eggs for a fixed duration, after which the adults were removed from the plate and the eggs were collected.  The egg-laying rate of animals at this stage is $\approx6$ eggs/animal/hour.  Synchronization could be tightened by shortening the duration of egg-laying, and the number of synchronized eggs could be increased by using more egg-laying adults.  

\paragraph{Nematode growth media - gelrite} NGM gelrite plates were made by replacing the Agar in normal NGM recipe with Gellan Gum (Sigma Aldritch), a polymer derived from algae. The recipe is given in Table \ref{NGM_Gelrite}.  In addition, peptone and cholesterol were omitted to prevent bacterial growth on the plate, so that the only available bacterial food was that which was initially inoculated.  Additive salts were prepared in 1 M stock solutions, and the $KH_2PO_4$ stock solution was adjusted to pH 6.
\begin{table}[htb]
\caption{NGM Gelrite}
\begin{center}
\begin{tabular}{|l|l|}
\hline
Ingredient & Amount \\
\hline
\multicolumn{2}{c}{Autoclave Together} \\
\midrule
NaCl & 3.0 g/l \\
Gellan Gum & 8.0 g/l \\
\hline
\multicolumn{2}{c}{Add Aseptically} \\
\midrule
$CaCl_2$ & 1 mM\\
$MgCl_2$ & 1 mM\\
$KH_2PO_4$ & 25 mM \\
\hline
\end{tabular}
\end{center}
\label{NGM_Gelrite}
\end{table}%

\paragraph{Phylogenetic tree of wild isolates} The phylogenetic tree of \textit{C. elegans} wild isolates was generated from the full CeNDR phylogenetic tree \cite{Cook2017-hq} which was hard-filtered by isotype using the \textit{prune} function in MATLAB.
\paragraph{Logistic fits}
Logistic fits were calculated using the MATLAB implementation of the Levenberg Marquardt \cite{Marquardt1963-ip} non-linear least squares algorithm within the \textit{lsqcurvefit} package. Logistic fits were performed on the raw data as well as the rescaled data. 
 The rescaled fits were nearly identical to the raw-fit parameters which were transformed according to Eq. \ref{eq:rescaled_logsitic} (See Supplemental Information Fig \ref{fig:rescaled_vs_raw}).

 \paragraph{Non-linear PCA} Nonlinear principal components analysis attempts to find an optimal auto encoder that can recreate the input data after passing it though a bottle neck layer with fewer components than the input.  The number of components in the bottle neck layer corresponds to the number of desired principal components.  The NLPCA implementation we use is from NLPCA toolbox for MATLAB \cite{Scholz2005-xp} and uses a multilayer perceptron architecture with a hyperbolic tangent activation function in the hidden layers. 
 
\paragraph{Linear Regression}  Each phenotypic output can be decomposed as a genetic contribution, an environmental contribution, and some residual error. Linear regression was performed using the \textit{fitlm} function in MATLAB.  if $y$ is the phenotypic variable of interest,
\[ y(G,E) = \beta_0 + \sum_i\beta_{1,i}G + \sum_j\beta_{2,j}E + \epsilon\]
were fit where $\epsilon$, with $y\in(\phi_1,\phi_2,\hat{l},\hat{r},\hat{A})$ , $G$ and $E$ are indicator variables which take the value of 1 or 0 depending of the strain $i$ and food $j$ in that condition, and $\epsilon$ are the residuals to be minimized. \textit{fitlm} uses an iteratively reweighted least squares algorithm.  For example, the best fit linear model for $\phi_1$ indicates that the average value of $\phi_1$ is $0.022$ and that changing to \textit{Bacillus pumilus} $0.115$ and switching to \textit{Pseudomonas fragi} moves $\phi_1$ $-0.200$.  F-statistics were calculated by the MATLAB \textit{anova} function, whichi si given by 
\[F = \mathbb{V}[\mathbb{E}(y_i)]/\mathbb{E}[\mathbb{V}(y_i)]\]
again, calculated for each prediction variable $y\in(\phi_1,\phi_2,\hat{l},\hat{r},\hat{A})$ and decomposed according to environment or genotype $i \in E,G$.  

\paragraph{Acknowledgements}
The authors would like to thank Marie-Anne Felix for providing us with wild-isolates of {\it C. elegans\/} as well as wild bacteria. We would also like to thank Stanislas Leibler, Bing Kan Xue, Maros Pleska, Jon Chuang, and Ricardo Rao for helpful discussion, and in particular S.L. for advice on the presentation of the material and B.K.X for suggesting the rescaling and the linear decomposition of the variance.

\printbibliography

\newpage
\onecolumn
\section*{Supplementary Materials}
\supplementarysection

\begin{figure*}[htp]
\begin{centering}
    \includegraphics[width=0.95 \textwidth]{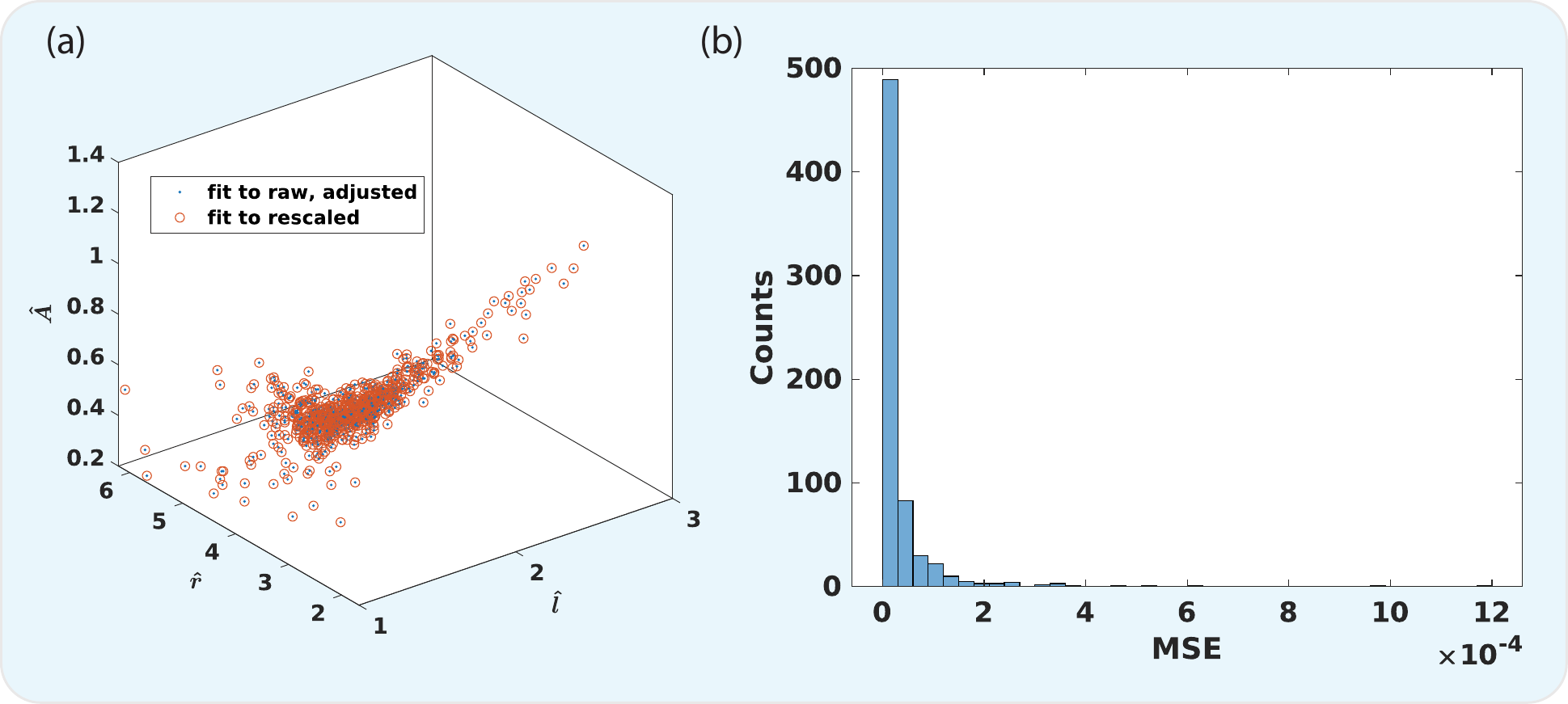}
    \caption{\textbf{(a)} Scatter plot of the three logistic fit parameters for each curve that were obtained either by fitting the raw data directly and re-normalizing the fit parameters with respect to the development time (blue points) or re-scaling the data and then performing the fit (orange points).  \textbf{(b)} Shows a histogram of the mean squared error between the raw adjusted and the re-sclaed fit data points.
    \label{fig:rescaled_vs_raw}
    }
    \end{centering}
\end{figure*}

\paragraph{Logistic fit rescaling}
Best fit logistic function parameters could be obtained from either the raw data or the rescaled data to yield equivalent results when the parameters fit from the raw data were adjusted according to Eq \ref{eq:rescaled_logsitic}. This can be seen in the 3-d scatterplot of the raw adjusted parameters (Fig. \ref{fig:rescaled_vs_raw}a), where the fit parameters for the rescaled data are shown as orange circles and the fit parameters from the raw data adjusted by Eq \ref{eq:rescaled_logsitic} are shows as blue dots.  If we quantify the mean squared error between the fits from these two different procedures, we find that most have 0 deviation and the maximum deviation is less than $2\cdot10^{-3}$ (Fig \ref{fig:rescaled_vs_raw}b).

\paragraph{Fraction on Food Analysis}
It has been noted that the development of \textit{C. elegans} in environments where they are completely surrounded by food is more rapid than on plates where they can be on or off food.  It is also known that the tendency for animals to be on food varies among strains \cite{Chang2006-vb,Bretscher2008-ce}.  In these experiments, animal behavior can have an influence on growth, as the experimental design provides an opportunity for the worms to leave and enter the food patch. Micro-fluidic chamber based systems are another alternative \cite{Uppaluri2015-nz} in which bacterial density can be precisely controlled, however, the aqueous environment can result in “thrashing” behavior, which has been shown to affect the animals physiology \cite{Laranjeiro2019-qe}. Because, in our systems, animals can freely move on and off food, it possible that behavioral differences could account for developmental changes. To asses this, we quantified whether animals that spend more time on food were more likely to develop more quickly.  We find that this does not seem to be the case, as there is no significant correlation between developmental duration and the fraction of time animals were found on food. In addition to individuals, the average development time was calculated for each combination of environment and genotype, and plotted against the total fraction of time spent by all individuals on food and this also does not show a significant correlation.  

\begin{figure*}[htp]
\begin{centering}
    \includegraphics[width=0.95 \textwidth]{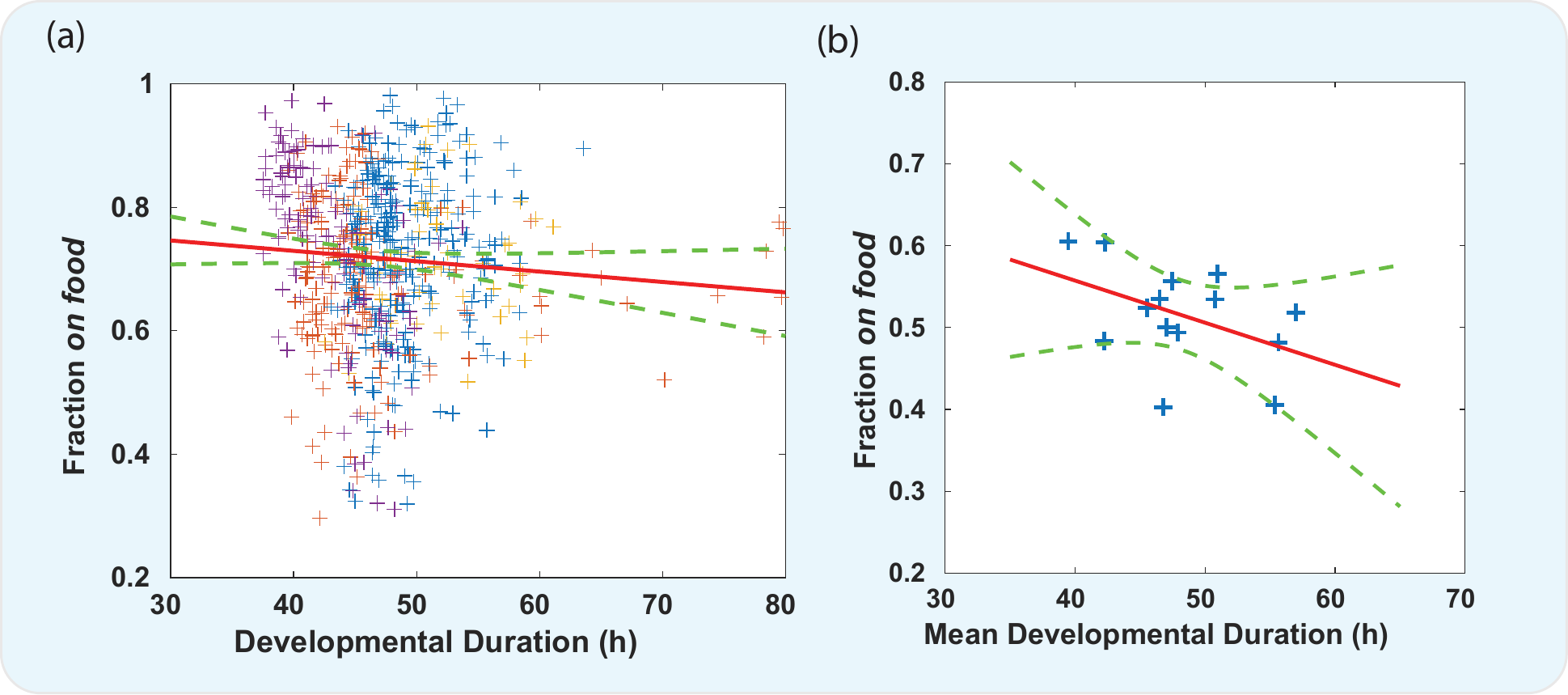}
    \caption{\textbf{(a)} For each individual, the fraction on food is plotted against that individuals developmental duration.  Fraction on food is the fraction of frames the animal id recorded less than 1.33mm from the center of the mini well. In addition, \textbf{(b)} shows the mean fraction on food and the mean developmental duration for all individuals in a condition. The best fit linear regression (red line) with is shown with its $95\%$ confidence interval (green dashed lines).  The slope of the regression is not statistically significantly different from 0 for either the individuals or the means in each condition. Colors represent the type of bacterial food for those individuals. \label{fig:fonf_correlations}
    }
    \end{centering}
\end{figure*}

\begin{figure*}[htp]
\begin{centering}
    \includegraphics[width=0.95 \textwidth]{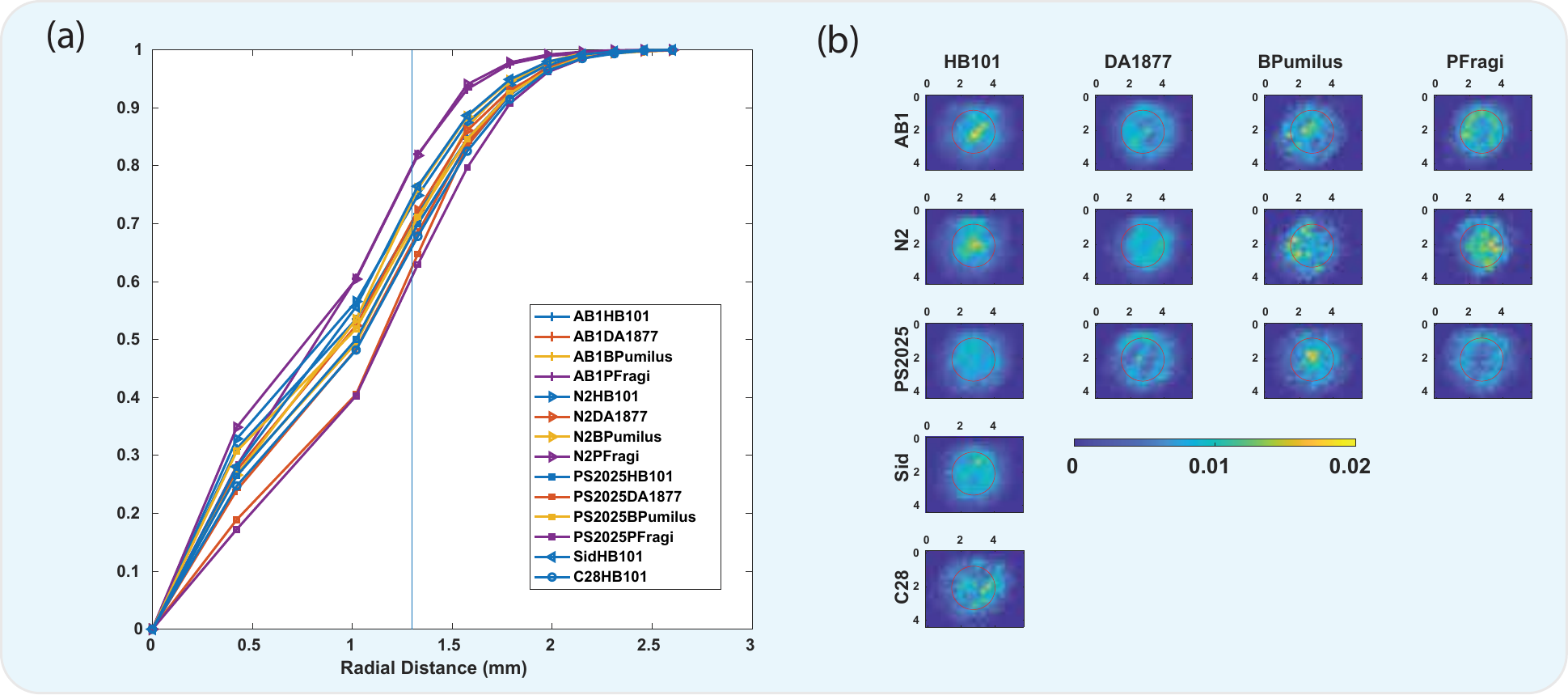}
    \caption{\textbf{(a)} Cumulative radial distribution functions of the centroid locations of all animals in a given condition as denoted by the combination of color (food) and symbol (genotype).  The blue line represents the demarcation between on food $(<1.33 mm)$ and off food $(\geq1.33 mm)$.  \textbf{(b)} Each heatmap shows the 2-d spatial distribution functions for all individuals in a given condition, estimated by a histogram.  Inside of the red circle indicates the on food region.
    \label{fig:fonf_dists}
    }
    \end{centering}
\end{figure*}

In this analysis, on food is taken to be less than 1.33 mm from the center of the well.  This roughly corresponds to the region on the mini plat where the 2 $\mu l$ of bacterial suspension is placed.  To calculate the on food percentage, the radial distance of each detected worm centroid is computed by subtracting the well center according to $r = \sqrt{(x-x_c)^2+(y-y_c)^2}$.  A radial distribution function (rdf) was then estimated with a histogram using equal area annular bins.  The overall (rdf) for each condition is shown in Fig \ref{fig:fonf_dists}(a) as a cumulative distribution function, in addition to the 2-d spatial distribution functions with the on food area denoted by a red circle.

\end{document}